\documentclass[aps,prl,preprint,groupedaddress]{revtex4}
\usepackage{graphicx}

\begin{document}



\title{Quantum phases in mixtures of fermionic atoms}

\author{C. Ates}
\affiliation{Max-Planck-Institut f\"ur Physik Komplexer Systeme\\
D-01187 Dresden, Germany}
\author{K. Ziegler}
\affiliation{Institut f\"ur Physik, Universit\"at Augsburg\\
D-86135 Augsburg, Germany}


\date{\today}

\begin{abstract}
A mixture of spin-polarized light and heavy fermionic atoms on a finite 
size 2D optical lattice is considered at various temperatures and values 
of the coupling between the two atomic species.
In the case, where the heavy atoms are immobile in comparison to
the light atoms, this system can be seen as a correlated binary 
alloy related to the Falicov-Kimball model. The heavy atoms represent
a scattering environment for the light atoms. The distributions of the 
binary alloy are discussed in terms of strong- and weak-coupling
expansions.
We further present numerical results for the intermediate interaction
regime and for the density of states of the light particles.
The numerical approach is based on a combination of a Monte-Carlo 
simulation and an exact diagonalization method.
We find that the scattering by the correlated heavy atoms can open a gap 
in the spectrum of the light atoms, either for strong interaction or small
temperatures.
\end{abstract}

\maketitle

\def\no{\noindent}

\section{Introduction}

Recent experimental progress in preparing and measuring clouds of ultracold
atoms in magnetic traps has opened a new way of studying bosonic and
fermionic many-particle quantum states. Among them are condensed and
Mott-insulating states of bosons in optical lattices (see e. g. \cite{jaksch98},
\cite{greiner02}, \cite{stoeferle04}, \cite{paredes04}).
In comparison with similar studies in solid-state physics, atomic clouds
enable us to design new many-particle systems by mixing different
types of atoms \cite{demarco99}, \cite{stamper00}, \cite{jochim03}. These mixtures can 
form new quantum states due to the competition between the different types of atoms. 
In this paper we propose a mixture
of light fermionic (e.g. $^6$Li) atoms and heavy fermionic (e.g. $^{40}$K) atoms, 
and study its low-temperature behavior in a (finite) optical lattice.
We assume that the cloud of this mixture is prepared in a magnetic
trap such that the atoms are spin polarized.

The difference of the masses of the two types of atoms implies two different
and well-separated time scales for their tunneling processes through the
optical lattice. The relatively fast tunneling processes of the light atoms
sets the relevant scale for the dynamics of the
mixture. In contrast, the relatively slow tunneling processes of the heavy atoms
are dynamically irrelevant and lead only to statistical fluctuations which drive
the system towards equilibrium. The latter will be discussed by the fact that the  
heavy particles form Ising-like (para-, ferro- and antiferromagnetic)
states. They provide a scattering environment for the light atoms. 
Formally, this physical
picture leads to the Falicov-Kimball model which has been used to describe complex
solid-state systems \cite{falicov69,freericks03,farkasovsky97,fradkin,gebhard}. 
A numerical study of the two-dimensional Falicov-Kimball model, based on exact
diagonalization, has revealed the possibility of a discontineous transition
between ordered and disordered phases \cite{farkasovsky97}. In the following we
will use analytic methods as well as a combination of exact diagonalization and 
Monte-Carlo
simulations to study large two-dimensional clusters for a better understanding of
the underlying physics.

The paper is organized as follows: In Sect. 2 the model is briefly discussed, 
based on a functional-integral representation. A mapping to a binary-alloy 
model is described in Sect. 3. This gives the foundation for our analytic 
treatment, based on weak- and strong-coupling expansions, and for the
construction of our numerical method. The numerical approach is used to
evaluate the distribution of the heavy atoms and the density of states of
the light atoms. 

\section{The Model}

The atomic degrees of freedom are given in second quantization by
local creation and annihilation operators. In the case of spin-polarized
fermionic atoms we use $c_{r}^\dagger$ ($c_{r}$) and $f_{r}^\dagger$ ($f_{r}$)
as the creation (annihilation) operators
for the light and the heavy fermionic atoms, respectively,
where $r$ denotes the coordinates of the site in the optical lattice.
The light
atoms can tunnel with tunneling rate ${\bar t}$, and we assume that the
tunneling rate of the heavy atoms is so small that we can neglect it.
Moreover, there is only a local interaction between the atoms in the
optical lattice, i.e. only atoms in the same potential well notice each
other. Since the atoms are spin-polarized fermions, there can be at most
one atom per sort in each potential well, thanks to Pauli's principle.
The interaction strength between light and heavy atoms is $U$.
This allows us to write the many-particle Hamiltonian as
\begin{equation}
H= -{\bar t}\sum_{\langle r,r'\rangle}c_r^\dagger
c_{r'}  + \sum_r\Big[-\mu (c_{r}^\dagger c_{r}+f_{r}^\dagger f_{r})
+Uf^\dagger_{r}f_{r}c^\dagger_{r}c_{r}\Big],
\label{hamilton}
\end{equation}
where $\langle r,r'\rangle$ means pairs of nearest-neighbor lattice sites.
We have assumed the same chemical potential $\mu$ for both types of atoms.
This may not be very general but will serve for the purpose of studying
competing quantum phases in the atomic mixture. The model defined in 
Eq. (\ref{hamilton}) is also known as the spinless Falicov-Kimball 
model \cite{falicov69,freericks03,farkasovsky97,fradkin,gebhard}. 
It is known to describe
ordered phases and phase transitions for correlated electronic systems
and was recently investigated intensively in the limit of infinite dimensions
\cite{freericks03}. We will study this model in the following for a finite
lattice, using a correlated binary-alloy (CBA) approach.

\subsection{Functional-Integral Representation}

A grand-canonical ensemble of a mixture of light and heavy fermionic atoms
at the inverse temperature $\beta=1/k_B T$ can be
defined by the partition function
\[
Z={\rm Tr} e^{-\beta H},
\]
where ${\rm Tr}$ is the trace with respect to all many-particle states
in the optical lattice. The Green's function
for the propagation of a light particle in the background formed 
by the heavy atoms in imaginary time $t$ is
\[
G(r,t;r',0)={1\over Z}{\rm Tr}\Big[
e^{-(\beta-t)H}c_re^{-tH}c_{r'}^\dagger
\Big].
\]
These expressions can also be written in terms of a functional
integral on a Grassmann algebra \cite{negele}. For the latter the
integration over a Grassmann field 
$\Psi_\sigma(r,t)$ and its conjugate ${\bar\Psi}_\sigma(r,t)$ ($\sigma=c,f$)
is given as a linear mapping from a Grassmann algebra to the complex
numbers. At a space-time point $(r,t)$ we have for integers $k,l\ge0$
\[
\int [{\bar\Psi}_\sigma(r,t)]^k[\Psi_\sigma(r,t)]^l
d\Psi_\sigma(r,t)d{\bar\Psi}_\sigma(r,t)=\delta_{k,1}\delta_{l,1}. 
\]
The partition function $Z$ of the grand-canonical ensemble then reads
\begin{equation}
Z=\int\exp(-S){\cal D}[\Psi_f,\Psi_c]
\label{funct1}
\end{equation}
with the action
\begin{equation}
S=\sum_{r,t,\sigma}{\bar\Psi}_\sigma(r,t)
[\Psi_\sigma(r,t)-\Psi_\sigma(r,t-\Delta)]+\Delta\sum_t
H[{\bar\Psi}_\sigma(r,t),\Psi_\sigma(r,t-\Delta)]
\label{action1}
\end{equation}
and the product measure
\[
{\cal D}[\Psi_f,\Psi_c]=\prod_{r,t,\sigma}
d\Psi_\sigma(r,t)d{\bar\Psi}_\sigma(r,t).
\]
The Green's function of light atoms is
\begin{equation}
G(r,t;r',t')=\langle\Psi_c(r,t){\bar\Psi}_c(r',t')\rangle.
\label{green}
\end{equation}
The discrete time is used with $t=\Delta, 2\Delta,...,\beta$, implying
that the limit $\Delta\to0$ has to be taken in the end and 
$\beta^{\prime}=\beta /\Delta$ is the number of time steps.
${\bar\Psi}_\sigma(r,t)$ and $\Psi_\sigma(r,t)$ are independent
Grassmann fields which satisfy antiperiodic boundary conditions in
time $\Psi_\sigma(r,\beta+\Delta)=-\Psi_\sigma(r,\Delta)$
and ${\bar\Psi}_\sigma(r,\beta+\Delta)=-{\bar\Psi}_\sigma(r,\Delta)$.
For the subsequent calculations it is convenient to rename 
$\Psi_\sigma(r,t)\to\Psi_\sigma(r,t+\Delta)$ because then the 
Grassmann field appears with the same time in the Hamiltonian of the
action (\ref{action1}).

\section{The Correlated Binary Alloy}

The functional integration in $Z$ and $G$ can be performed in several steps,
beginning with the integration of the heavy atomic field $\Psi_f$,
introducing the Ising spins and finally integrating the light atomic
field $\Psi_c$ \cite{ziegler02}. The details of this procedure are described
in Appendix A. As a result we obtain for the partition function
\[
Z=\sum_{\{S(r)\}}Z(\{S_r\})
\]
with
\begin{equation}
Z(\{S_r\})=
{\bar\mu}^{\beta'\sum_r(1+S(r))/2}
{\rm det}(-\partial_t+{\bar\mu}+{\hat t}-(U'/2{\bar\mu})(1+S)).
\end{equation}
The parameters are $U'=\Delta U$, ${\bar \mu}=1+\Delta \mu$,
and ${\hat t}$ is the
tunneling term multiplied by $\Delta$. Moreover, $\partial_t$ is the
time-shift operator. 
The Ising spin $S(r)$ corresponds with a local occupation number
$n_f(r)$ of the heavy atoms as
\[
n_f(r)=[1+S(r)]/2.
\]
The Green's function is now an averaged resolvent
\begin{equation}
G=\langle (-\partial_t+{\bar\mu}+{\hat t}-(U'/2{\bar\mu})(1+S))^{-1}
\rangle_{\rm Ising},
\label{green1}
\end{equation}
where the average $\langle ... \rangle_{\rm Ising}$ is taken with
respect to the distribution
\begin{equation}
P(\{ S(r)\})={Z(\{S_r\})\over \sum_{\{S(r)\}}Z(\{S_r\})}.
\label{distr11}
\end{equation}
The partition function can also be written as
\begin{equation}
Z=\sum_{\{S(r)\}}{\bar\mu}^{\beta'\sum_r(1+S(r))/2}
{\rm det}[{\bf 1}+\{{\bar\mu}+{\hat t}-(U'/2{\bar\mu})(1+S)\}^{\beta'}].
\label{partf}
\end{equation}
The distribution is not $Z_2$ invariant (i.e. invariant under a change 
$S(r)\to-S(r)$), except for a half-filled lattice (i.e. $\mu=U/2$).

\noindent
The representation of the Green's function in Eq. (\ref{green1}) 
is our main analytic result. It
means that the light particles tunnel through the optical lattice where
they are scattered by the heavy particles. The distribution of the
heavy particles is given by the distribution shown in Eq. (\ref{distr11}).
The latter depends on the temperature but also on the parameters of
the light particles like ${\bar t}$ and the coupling $U$ between the
light and the heavy particles. This reflects the intimate relationship
between the two types of particles. In other words, the light particles
move in a random potential formed by the heavy particles. This randomness,
formally expressed by the Ising spins, is correlated and can be called
correlated binary alloy (CBA). There is a
correlation length which diverges at the phase transitions of the
Ising system. In the subsequent investigation we will study these
phases and their implications for the properties of the light particles. 

The symmetric matrix ${\bar\mu}-{\hat t}+(U'/2{\bar\mu})(1+S)$ can be
diagonalized with eigenvalues $1-\Delta\lambda_j$. Then the determinant
in Eq. (\ref{partf}) is for
$\Delta\sim0$
\[
{\rm det}[{\bf 1}+\{{\bar\mu}+{\hat t}-(U'/2{\bar\mu})(1+S)\}^{\beta'}]
\sim\prod_j(1+e^{-\beta\lambda_j}).
\]
Since the matrix depends on the fluctuating Ising spins, it is difficult to
determine the eigenvalues. One way to get an idea about the physics
of this model is to study the asymptotic regimes of strong and weak
coupling, another one is to use a numerical diagonalization procedure.
Both approaches shall be applied subsequently. 


\subsection{Approximations of the CBA Distribution}

The distribution was studied in the case of strong coupling 
(i.e. the tunneling 
(or $U^{-1}$) expansion) in a number of papers \cite{gruber95,ziegler02}. It 
leads at half-filling to an Ising model with $Z_2$ symmetry. In the following 
we study the CBA distribution in weak and strong-coupling approximations as
well as numerically by a Monte-Carlo simulation. The density of states of
the light atoms are evaluated by a numerical procedure.

\subsubsection{System without Tunneling}

The absence of the $Z_2$-symmetry can
be observed already in the absence of tunneling. Then we have in the limit 
$\Delta\to0$
\[
P_0(\{ S(r)\})=\prod_r
{e^{\beta\mu(1+S(r))/2}+e^{\beta[\mu+(\mu-U)(1+S(r))/2]}
\over 1+2e^{\beta\mu}+e^{\beta(2\mu-U)}
}
\]
which is $Z_2$ invariant  
only for $\mu=U/2$. 
\begin{figure} 
\begin{center}
\scalebox{0.14}
{\includegraphics[scale=3]{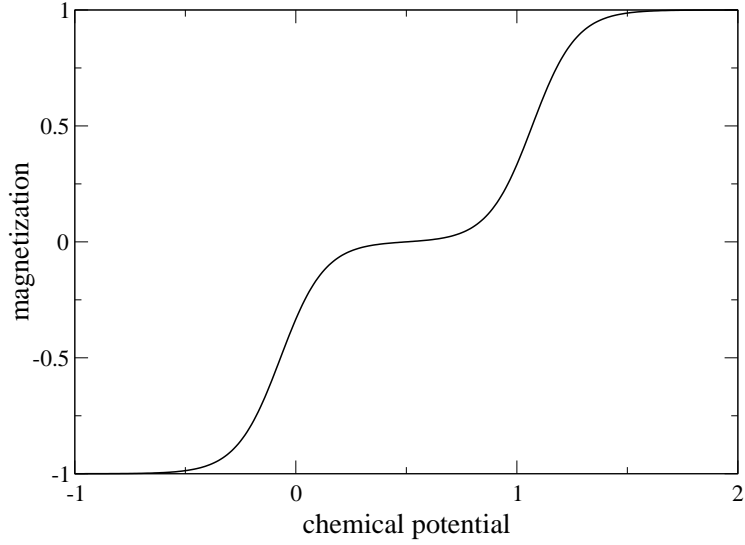}}
\end{center}
\caption{Average spin for a system without tunneling: $U=1$ and $\beta=10$.}
\label{aspin}
\end{figure}
The average spin is shown in Fig. \ref{aspin} and its asymptotic
low-temperature behavior is
\[
\langle S\rangle=
{e^{\beta(2\mu-U)}-1\over 1+2e^{\beta\mu}+e^{\beta(2\mu-U)}}\sim
\cases{
-1 & $\mu<0$ \cr
0  & $0\le \mu <U$ \cr
1/3 & $\mu=U$ \cr
1  & $U<\mu$ \cr
}. 
\]
Thus only for $0<\mu <U$ the Ising state is paramagnetic.
The other regimes are ferromagnetic. In terms of the configurations
of the heavy atoms there are no heavy atoms for $\mu<0$ and a
heavy atom at each site for $\mu>U$. In the intermediate regime
$0<\mu <U$ we anticipate that the coupling of neighboring Ising spins,
caused by a non-zero tunneling rate ${\bar t}$, will lead to ordered
Ising spins at low temperatures. 

\subsubsection{Tunneling Expansion}

The effect of a weak tunneling rate can be evaluated in terms of
a perturbation theory with respect to ${\bar t}$. Moreover, we
take the limit $\Delta\to0$ and consider the asymptotic regime of
low temperatures (i.e. $\beta\sim\infty$). If we include
tunneling terms up to order $O({\bar t}^2)$ we get for $\mu=U/2$ 
the Ising model with nearest-neighbor spin interaction:
\begin{equation}
P_s(\{ S(r)\})={
\exp\Big[-\beta{{\bar t}^2\over 2U}
\sum_{\langle r,r'\rangle}S(r)S({r'})+o({\bar t}^3)\Big]
\over
\sum_{\{S(r)=\pm1 \}} \exp\Big[-\beta{{\bar t}^2\over 2U}
\sum_{\langle r,r'\rangle}S(r)S({r'})+o({\bar t}^3
)\Big]
}.
\label{distr2}
\end{equation}
This model has an antiferromagnetic low-temperature phase.

\no
The spin-spin coupling is exponentially small in $\beta$ for
$\mu<0$ and $\mu>U$ but of order ${\bar t}^2/U$ for $0<\mu <U$.
In particular, we can distinguish three different regimes:
\[
\mu<0:\ \ \ 
P_s(\{ S\})\propto\exp\Big[-\beta{|\mu|\over 2}\sum_r S(r)
+o({\bar t}^3)\Big],
\]
\[
0<\mu<U:\ \ \ 
P_s(\{ S\}) \propto
\exp\Big[
{\beta^2{\bar t}^2\over8}e^{-\beta U/2}\sinh[\beta(\mu-U/2)]
\sum_rS(r)
-\beta{{\bar t}^2\over4U}
\sum_{<r,r'>}S(r)S(r')+o({\bar t}^3)\Big]
\]
and
\[
U<\mu:\ \ \ 
P_s(\{ S\})\propto\exp\Big[\beta{\mu-U\over 2}\sum_r S(r)+o({\bar t}^3)\Big].
\]
Besides the two ferromagnetic regimes for $\mu<0$ and $\mu>U$
we have the intermediate regime $0<\mu<U$ with antiferromagnetic ordering. 
There is an exponentially small magnetic field for $\mu\ne U/2$ which 
breaks the $Z_2$ symmetry. As we approach $\mu=0$ or $\mu=U$ the magnetic
field becomes larger. There is a first order transition from the antiferro- to
a ferromagnetic phase when the magnetic field starts to dominate the
spin-spin interaction. This can be seen in a simple mean-field approximation.

\subsubsection{The Weak-coupling Limit}

In the case of weak coupling we can perform an expansion in terms of
the coupling parameter $U$. This gives in leading order an uncorrelated 
binary alloy:
\[
P_w(\{ S(r)\})=\prod_r
{e^{\beta(\mu-Ug)S(r)/2}
\over \sum_{S(r)=\pm1}e^{\beta(\mu-Ug)S(r)/2},
}
\]
where $\epsilon(k)$ is the dispersion of the tunneling term and
\[
g(\mu)=\int\Theta(\epsilon(k)+\mu){d^dk\over (2\pi)^d},
\]
i.e. $0\le g\le 1$.
Thus the Ising groundstate for weak coupling is ferromagnetic.
In particular,
\[
\langle S\rangle =\tanh (\beta(\mu-Ug(\mu))/2).
\]

\subsection{The Density of States}

\begin{figure}
\centering
\includegraphics[width=0.4\textwidth]{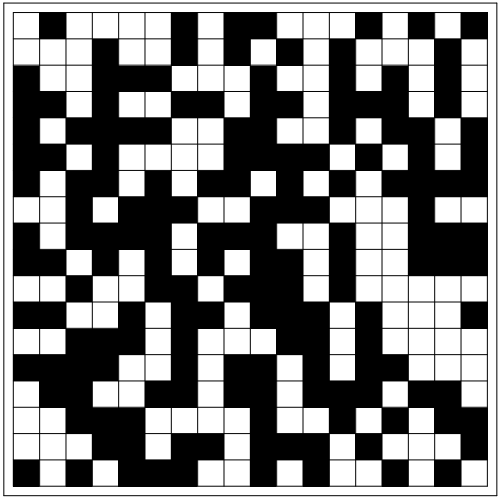}
\caption{Paramagnetic Ising-spin configuration for 
${\bar t}=1$, $U=3\, ,\; \mu=U/2$, and $\beta=3$. White (black) 
squares refer to sites (un)occupied with a heavy atom.}
\label{config1}
\end{figure}

\begin{figure}
\centering
\includegraphics[width=0.4\textwidth]{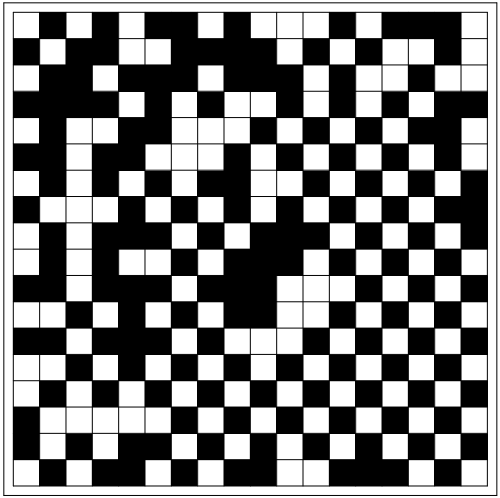}
\caption{Mixture of para- and antiferromagnetic Ising-spin textures
for ${\bar t}=1$, $U=3\, ,\; \mu=U/2$, and $\beta=7$} 
\label{config2}
\end{figure}

\begin{figure}
\centering
\includegraphics[width=0.4\textwidth]{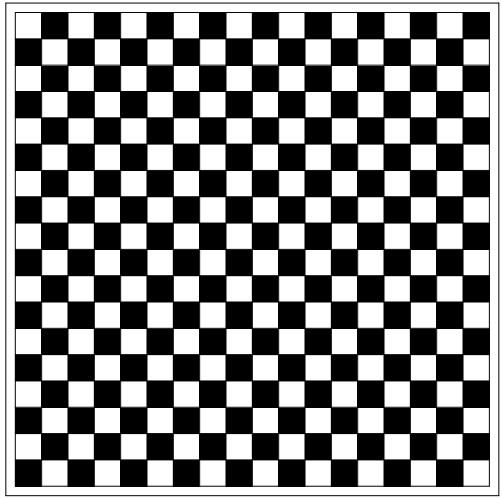}
\caption{Antiferromagnetic Ising-spin configuration for ${\bar t}=1$, $U=3\, ,\; \mu=U/2$, 
and $\beta=14$}
\label{config3}
\end{figure}

\begin{figure}
\centering
\includegraphics[width=0.4\textwidth]{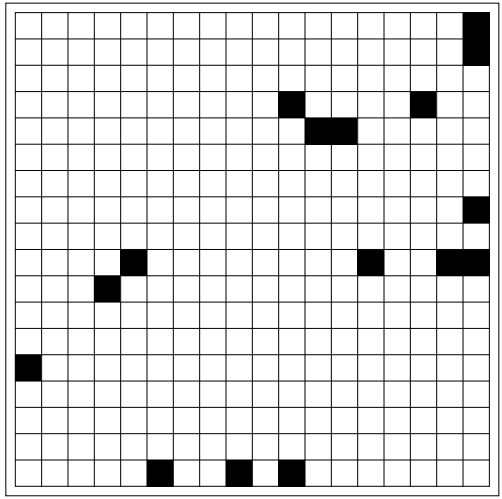}
\caption{Ferromagnetic Ising-spin configuration for ${\bar t}=1$, $U=8\, ,\; \mu=0.8U$, 
and $\beta=14$} 
\label{config4}
\end{figure}

The density of states (DOS) for the light particles can be obtained from the
diagonal elements of the Green's function in Eq. (\ref{green1}). Its
qualitative behavior depends strongly on the state of the heavy
particles: In the case of a ferromagnetic state the DOS shows a
single band, for the antiferromagnetic state it has a gap. 
For a paramagnetic state of the heavy particles the form of the DOS is 
less obvious. We have calculated the DOS numerically for a $18\times 18$
square lattice with open boundaries. For this purpose we have generated
configurations of the Ising spins according to the distribution function 
in Eq. (\ref{distr11}), using the Metropolis algorithm.
Typical configurations with large statistical weight are shown in Figs. \ref{config1}
- \ref{config4} for different values of the physical parameters $U$, $\mu$,
and $\beta$.

For a given configuration $\{ S(r)\}$ of Ising spins the Hamiltonian
\[
h={\hat t}-{U\over 2}S
\]
is diagonalized. From the eigenvalues $\lambda_k(\{ S(r)\})$ the DOS 
of $h$ is calculated as
\begin{equation}
D(E,\{ S(r)\}) = \frac{1}{N\pi}\sum_{k=1}^N \delta\left(E 
- \lambda_k(\{ S(r)\})\right) \quad ,
\end{equation}
where $N$ is the number of lattice sites. Finally, the DOS related to
the Green's function in Eq. (\ref{green1}) 
is determined by averaging over $L=100$ spin configurations:
\begin{equation}
D(E)=\frac{1}{L}\sum_{\{ S(r)\}} D(E,\{ S(r)\}) \quad .
\end{equation}
In the following the hopping rate is set to ${\bar t}=1$.

\begin{figure}
\centering
\includegraphics[width=0.8\textwidth]{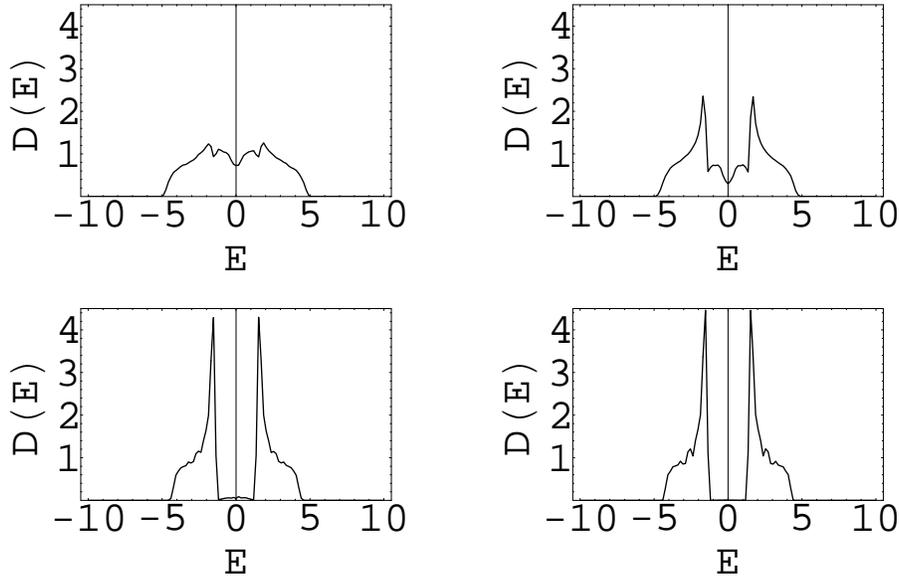}
\caption{DOS for $U=3\, ,\; \mu=U/2$ (half filling) and different temperatures.
First row: $\beta=3 \, ,\; \beta=7\, .$ Second row: $\beta=10\, ,\; 
\beta=14\,.$} \label{DOS1}
\end{figure}

\noindent
Fig. \ref{DOS1} shows the DOS of the light particles for $U=3$ and half 
filling (i. e. $\mu=U/2$) at different temperatures.
For small $\beta$, i. e.  high temperatures, the system shows a gapless
metallic band, which is symmetric around the Fermi level, and the Ising 
spins form a paramagnetic state. The DOS is slightly suppressed at the band 
center due to the interaction between the light and the heavy atoms.
When the temperature is decreased, the Ising spins start to order
antiferromagnetically, i. e. the heavy atoms create a chessboard-like phase
with empty sites.
This is accompanied by the formation of a gap around the Fermi level and a
strong enhancement of the DOS at the inner band edges. Very similar results
were found in a dynamical cluster approximation on an $8\times8$ cluster 
\cite{hettler00}.

\begin{figure}
\centering
\includegraphics[width=0.8\textwidth]{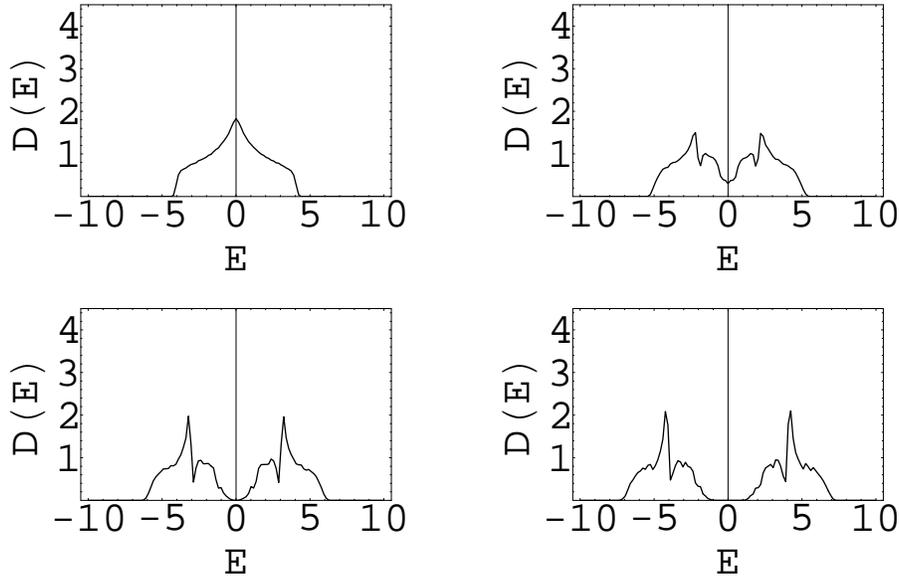}
\caption{DOS for $\beta=3$ at various values of the interaction $U$ and
half-filling ($\mu=U/2$). First row: $U=1\, ,\;U=4, .$ Second row: $U=6\,
,\; U=8\, .$} \label{DOS2}
\end{figure}

\noindent
The high-temperature regime of the system at half filling is depicted in Fig.
\ref{DOS2} for various interaction strengths $U$.
For small interaction the DOS shows a metallic band and is peaked around the
Fermi level. 
For increasing $U$ this peak gets suppressed and a band splitting to two 
symmetric bands occurs.
The spectral weight within these subbands is highest at their center.
A further increase of $U$ leads to a shift of the lower and upper band to lower
and higher energies, respectively.

\begin{figure}
\centering
\includegraphics[width=0.8\textwidth]{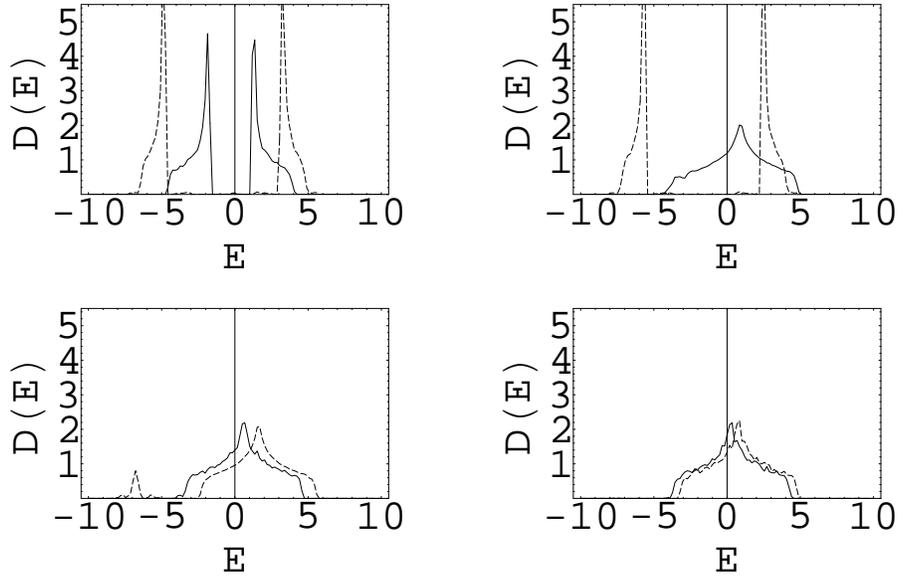}
\caption{DOS for $\;\beta=14\, ,$ $U=3$ (solid), $U=8$ (dashed) and 
different values of the chemical
potential. First row: $\mu=0.6U\, ,\;\mu=0.7U\, .$ Second row: $\mu=0.8U\,
,\;\mu=0.9U\, .$} \label{DOS3}
\end{figure}

\noindent
Figure \ref{DOS3} shows the DOS in the low-temperature regime for two
values of the interaction strength ($U=3$, solid and $U=8$, dashed) 
and different values of the chemical potential.
For the latter the distribution (\ref{distr11}) has no $Z_2$ symmetry,
i.e. is not invariant under a global spin flip $S\to -S$.
Near half filling the heavy atoms order in a chessboard configuration. 
This behavior is stablized for larger deviations fom $\mu=U/2$ when the
interaction strength is increased. As $\mu$ deviates even further from $U/2$ 
the Ising spins start to order ferromagnetically.
The spectral weight locally shifts to the center of each subband and globally
shifts from the lower to the upper band. 
For the completely ordered Ising spins the lower band disappears and the 
system has only one band.


\section{Conclusions}

A mixture of light and heavy fermionic atoms is studied as a system 
in which the light atoms live in a correlated disordered environment.
This environment is formed by the heavy atoms.
The disorder is given by fluctuating Ising spins with a complex 
temperature-dependent distribution.
This distribution is $Z_2$ (spin-flip) invariant only at half filling 
(i.e. $\mu=U/2$) but has a broken spin-flip symmetry for $\mu\ne U/2$.
This symmetry breaking favors an Ising spin $S_r=-1$ at low density (i.e. $\mu<0$)
and $S_r=1$ at high density (i.e. $\mu>U$). There is an intermediate regime
where an antiferromagentic (staggered) Ising-spin configuration is favored.
Scattering on these configurations opens a gap in the band of the light atoms.
\vskip0.5cm
\noindent
Acknowledgement:
\vskip0.2cm
\noindent
This work was supported by the Sonderforschungsbereich 484.

\section*{A. Appendix}
\subsection*{A.1. Integration of the Heavy Atoms}

It is possible to integrate out the field $\Psi_f$ of the heavy
atoms in Eqs. (\ref{funct1}) and (\ref{green}), since it 
appears in $S$ only as a quadratic form:
\[
S=S_c+S_f+S_I
\]
with
\[
S_c=\sum_t\Big\{\sum_r
\big[{\bar\Psi}_c(r,t)\Psi_c(r,t+\Delta)
-{\bar\mu}{\bar\Psi}_c(r,t)\Psi_c(r,t)\big]
-\tau\sum_{\langle r,r'\rangle}{\bar\Psi}_c(r,t)
\Psi_c(r',t)\Big\}
\]
\[
S_f=\sum_t\Big\{\sum_r
\big[{\bar\Psi}_f(r,t)\Psi_f(r,t+\Delta)
-{\bar\mu}{\bar\Psi}_f(r,t)\Psi_f(r,t)\big]
\Big\}.
\]
The interaction between the two types of atoms is given by
\[
S_I=U'\sum_{r,t}
{\bar\Psi}_f(r,t)\Psi_f(r,t)
{\bar\Psi}_c(r,t)\Psi_c(r,t).
\]
The integration over the Grassmann field $\Psi_f$ in $Z$ gives a
space-diagonal determinant
\begin{equation}
\int e^{-S_f-S_I}\prod_{r,t}
d\Psi_f(r,t)d{\bar\Psi}_f(r,t)
={\rm det}(-\partial_t+{\bar\mu} -U'{\bar\Psi}_c\Psi_c),
\label{det}
\end{equation}
where $\partial_t$ is the time-shift operator
\[
\partial_t\Psi(r,t)=\cases{
\Psi(r,t+\Delta) & $\Delta\le t < \beta$ \cr
-\Psi(r,\Delta) & $t=\beta$ \cr
}.
\]
The second equation is a consequence of the antiperiodic boundary
condition of the Grassmann field. 

\subsection*{A.2. Expansion with Ising Spins}

The partition function is now a functional integral of the 
$c$--Grassmann field
\[
Z=\int e^{-S_c} {\rm det}(-\partial_t+{\bar\mu}
-U'{\bar\Psi}_c\Psi_c)
{\cal D}[\Psi_c]
=\int e^{-S_c}\prod_r\Big[1+\prod_t
({\bar\mu}-U'{\bar\Psi}_c(r,t)
\Psi_c(r,t))\Big]
{\cal D}[\Psi_c].
\]
The product can be expanded in terms of Ising spins $\{ S(r)=\pm 1\}$ 
\cite{ziegler02} as
\[
\prod_r\Big[1+\prod_t
({\bar\mu}-U'{\bar\Psi}_c(r,t)
\Psi_c(r,t))\Big]
=\sum_{\{S(r)=\pm1\}}\prod_r
\prod_t[{\bar\mu}-U'{\bar\Psi}_c(r,t)\Psi_c(r,t)
]^{1+S(r)\over 2}.
\]
This reads with ${\bf I}={1+S\over 2}$ as
\[
=\sum_{\{S(r)=\pm1\}}
\prod_r{\bar\mu}^{\beta'{\bf I}(r)}e^{-(U'/{\bar\mu})
{\bf I}(r)\sum_t{\bar\Psi}_c(r,t)\Psi_c(r,t)}.
\]
Now the partition function $Z$ can be expressed by a summation over
configurations of Ising spins as
\[
Z=\sum_{\{S(r)=\pm1\}}Z(\{S(r)\})
\]
with
\begin{equation}
Z(\{S(r)\})=\int e^{-S_c}
\prod_r{\bar\mu}^{\beta'{\bf I}(r)}e^{-(U'/{\bar\mu})
{\bf I}(r)\sum_t{\bar\Psi}_c(r,t)\Psi_c(r,t)}
{\cal D}[\Psi_c].
\label{a}
\end{equation}

\subsection*{A.3. Integration of the Light Atoms}

The $c$--Grassmann field appears only in a quadratic form
in the partition function:
\[
S_c'=S_c+{U'\over{\bar\mu}}
\sum_{r,t}{\bf I}(r){\bar\Psi}_c(r,t)\Psi_c(r,t)
\equiv{\bar\Psi}_c\cdot(\partial_t-{\bar\mu}-{\hat t}+
{U'\over{\bar\mu}}{\bf I})\Psi_c.
\]
After performing the $\Psi_c$--integration we obtain
\[
Z(\{S_r\})={\bar\mu}^{\beta'\sum_r{\bf I}(r)}
{\rm det}(-\partial_t+{\bar\mu}+{\hat t}-{U'\over{\bar\mu}}{\bf I}).
\]
Following the same procedure for the Green's function, we obtain
for $G$ in Eq. (\ref{green}) the matrix
\[
G={\sum_{\{S(r)\}}
(-\partial_t+{\bar\mu}+{\hat t}-(U'/{\bar\mu}){\bf I})^{-1}
{\bar\mu}^{\beta'\sum_r{\bf I}(r)}
{\rm det}(-\partial_t+{\bar\mu}+{\hat t}-{U'\over{\bar\mu}}{\bf I})
\over\sum_{\{S(r)\}}{\bar\mu}^{\beta'\sum_r{\bf I}(r)}
{\rm det}(-\partial_t+{\bar\mu}+{\hat t}-{U'\over{\bar\mu}}{\bf I})}.
\]

\end{document}